\documentstyle[11pt]{article}
\DeclareFontFamily{OT1}{msb}{}{}
\DeclareFontShape{OT1}{msb}{m}{n}
 {  <5> <6> <7> <8> <9> <10> gen * msbm
      <10.95><12><14.4><17.28><20.74><24.88>msbm10}{}
\DeclareMathAlphabet{\bubble}{OT1}{msb}{m}{n}

\begin{document}
\rightline{hep-th/9806243}
\begin{center}
{\Large  Lorentz covariance, higher-spin superspaces and self-duality}
\footnote{Presented by Jean Nuyts at the International Conference on
Particles, Fields and Gravitation, \L{}odz, Poland (April 15-19 1998); 
to be published in the Proceedings.}
\vskip 0.4 true cm
{\large Chandrashekar Devchand$^a$\ \ and\ \  Jean Nuyts$^b$}
\vskip 0.3 true cm
{\small
{devchand@mis.mpg.de, nuyts@umh.ac.be}\\[5pt]
{\it $^a$ Max-Planck-Institut f\"ur Mathematik in den Naturwissenschaften}\\
{\it Inselstra\ss{}e 22-26, 04103 Leipzig, Germany}\\[5pt]
{\it $^b$  Physique Th\'eorique et Math\'ematique, 
Universit\'e de Mons-Hainaut}\\
{\it 20 Place du Parc, 7000 Mons, Belgium}}
\end{center} 
\vskip 0.5 true cm
\begin{abstract} 
\noindent
Lorentz covariant generalisations of the notions of supersymmetry, superspace
and self-duality are discussed.  The essential idea is to extend standard
constructions by allowing tangent vectors and coordinates which transform
according to more general Lorentz representations than solely the spinorial and
vectorial ones of standard lore.  Such superspaces provide model configuration
spaces for theories of arbitrary spin fields.  Our framework is an elegant one
for handling higher-dimensional theories in a manifestly SO(3,1) covariant
fashion.  A further application is the construction of a hierarchy of solvable
Lorentz covariant systems generalising four-dimensional self-duality.
\end{abstract}

\renewcommand{\Large}{\large}
\section{Introduction} 

There exist various forms of higher dimensional fundamental building blocks for
the observed four dimensional `real world'.  An important question is the
recovery of four-dimensional Lorentz invariance, the most important observed
symmetry in physics.  It often seems desirable to have a framework in which the
higher dimensional objects transform covariantly under $SO(3,1)$ and therefore
have some intrinsic four-dimensional nature.  Recent developments in string
physics, moreover, seem to require an extension of the standard notions of
supersymmetry and superspace.  Over the past couple of years, we have been
investigating $SO(4,{\bubble C})$-covariant superalgebras which provide 
generalisations of supersymmetry and a manifestly Lorentz-covariant framework 
for the investigation of spaces of dimensions greater than four.

Standard superspace is a homogeneous space constructed as the quotient of the
super Poincar\'e group by the Lorentz group, with four bosonic coordinates 
$\ \{ Y^{\alpha{\dot\alpha}} \}$ and four fermionic coordinates 
$\ \{Y^\alpha, Y^{\dot\alpha} \}$, transforming according to the vectorial 
$({\textstyle{1\over 2}},{\textstyle{1\over 2}})$ and spinorial 
$(0,{\textstyle{1\over 2}})$ and $({\textstyle{1\over 2}},0)$ representations
of the Lorentz group.  Superspace can therefore be considered as a direct sum 
of 4d Minkowski space with two 2d spinorial spaces.  Now, once the 
restriction to coordinates transforming solely according to the vectorial 
$({\textstyle{1\over 2}},{\textstyle{1\over 2}})$ representation
has been abandoned, one can ask whether one can build more general superspaces
from direct sums of more general sets of representation spaces than the usual
spinorial and vectorial ones; namely, spaces with coordinates from the set of
general Lorentz tensors $\
\{Y^{\alpha_1\dots\alpha_{2s}{\dot\alpha}_1\dots{\dot\alpha}_{2{\dot s}}}\}$,
symmetrical in the $\alpha$ indices and in the ${\dot\alpha}$ indices,
for some collection of representations $\ \Lambda_p = \{ (s,{\dot{s}}) \}\ $,
where $p$ denotes the maximum value of $\,s{+}{\dot{s}}\,$ occurring in the
chosen set of representations.

This is clearly a way of parametrising a higher-dimensional space in a
manifestly 4d Lorentz covariant fashion.  Usually, in higher-dimensional
theories, the recovery of 4d Lorentz covariance is achieved by having the
extra coordinates transform as a bunch of Lorentz scalars.  The idea here is
to start with a set of Lorentz representations and hence fix the 4d Lorentz
structure to the coordinate system.  For instance a simple bosonic extension
of 4d space with coordinates 
$\ \{Y^{\alpha{\dot\alpha}},Y^{\alpha\beta\gamma{\dot\alpha}} \}$ has 
dimension $4{+}8{=}12$.  For the sake of generality, we consider a graded 
vector space with the coordinates $\{Y(s,{\dot{s}})\}$ spanning a 
supercommutative
${\bubble Z}_2$-graded algebra, $\,{\cal V}{=}{\cal V}_0 {+}{\cal V}_1\,$.  
${\cal V}_0 $ (resp.  ${\cal V}_1$) contains bosonic (resp. fermionic) 
coordinates if $2(s{+}{\dot{s}})$ is even (resp.  odd).  Each representation 
$\,Y(s,{\dot{s}})\,$ included, increases the bosonic (resp.  fermionic)
dimension of the hyperspace ${\cal M}$\ by $(2s{+}1)(2{\dot{s}}{+}1)$.  

\section{Spin-p Heisenberg superalgebras}
To describe such hyperspaces as the homogeneous spaces of some algebra, the
super-Poincar\'e algebra needs to be extended to some algebra including
higher-spin generators.  Such enhancement of customary supersymmetry, going
beyond the Haag-\L{}opusanski-Sohnius barrier, has been extensively studied by
Fradkin and Vasiliev \cite{FV}.  These authors were motivated by physical
considerations to realise such higher-spin algebras on 4d de Sitter space
fields.  Consistency of the dynamics required the inclusion of {\it all} 
spins, yielding infinite dimensional algebras realised on infinite chains 
of fields having spins all the way up to infinity. 

Our approach \cite{DN1,DN2,DN3} has been more abstract and does {\it not}
require infinite dimensionality of the algebra. We consider some basically
finite set of higher-spin generators (in addition to the spin $1$ and spin 
${\textstyle{1\over 2}}$ super-Poincar\'e generators); and we interpret the
higher-spin generators 
$X_{\alpha_1\dots\alpha_{2s}{\dot\alpha}_1\dots{\dot\alpha}_{2{\dot s}}}$ 
as `momenta' in extra dimensions parametrised by
higher-spin coordinates. We do not make any a priori field theoretical or
dynamical requirements; and we realise our algebras in flat space. 
The grading, however, remains a ${\bubble Z}_2$ one, with 
all integer-spin representations in an even-statistics (bosonic) subspace
${\cal A}_0$ and all half-integer-spin representations in an odd-statistics
(fermionic) subspace ${\cal A}_1$. Thus the Heisenberg superalgebra associated 
to standard (complexified) superspace 
with $\ \Lambda_1 = \{ ({\textstyle{1\over 2}},{\textstyle{1\over 2}}), 
({\textstyle{1\over 2}}, 0), (0,{\textstyle{1\over 2}}) \}\ $ and
non-zero canonical supercommutation relations,
$$
\begin{array}{llll}
&\left\{  X_{\alpha} , X_{{\dot\beta}}\right\}  =\ X_{\alpha{\dot\beta}}\  ,
\quad
&\left[  X_{\alpha} , Y^{\beta{\dot\beta}} \right]    = \  
\delta^\beta_\alpha Y^{\dot\beta} \  ,
\quad
&\left[  X_{{\dot\alpha}}, Y^{\beta{\dot\beta}}\right]   =    
\delta^{\dot\beta}_{\dot\alpha} Y^{\beta}
\\[5pt]
&\left[  X_{\alpha{\dot\alpha}}, Y^{\beta{\dot\beta}} \right]    =  
\delta^\beta_\alpha  \delta^{\dot\beta}_{\dot\alpha}\  ,
&\left\{  X_{\alpha} , Y^{\beta } \right\}   =   \delta^\beta_\alpha\  ,
&\left\{  X_{{\dot\alpha}}, Y^{\dot\beta}  \right\}   =  
\delta^{\dot\beta}_{\dot\alpha}\  \  ,
\end{array}
$$
is generalised to a {\it spin p Heisenberg superalgebra} 
$\ {\cal G}{=}{\cal A}{+}{\cal V}\ $, based on a set of representations 
$\ \Lambda_p = \{ (s,{\dot{s}}) \}\ $ appearing in $\,{\cal A}{+}{\cal V}\,$. 

For any specific $(s,{\dot{s}})\in \Lambda_p$, we label the 
$(2s{+}1)(2{\dot{s}}{+}1)$ 
components of the coordinate tensor $\,Y(s,{\dot{s}})$ as 
$\,Y(s,s_3\,;{\dot{s}},{\dot{s}}_3) $, where $s_3$ (resp.  ${\dot{s}}_3$) 
run from $-s$ to $s$ (resp.  from $-{\dot{s}}$ to ${\dot{s}}$) in integer 
steps. These are in one-to-one correspondence with the components
in standard 2-spinor index notation, 
$\ Y^{\alpha_1\dots\alpha_{2s}{\dot\alpha}_1\dots{\dot\alpha}_{2{\dot{s}}}}$.
The span of coordinates $\{ Y(s,s_3\,;{\dot{s}},{\dot{s}}_3)\}$, 
for all $(s,{\dot{s}})$ in the chosen set $\Lambda_p$, is defined to be 
a supercommutative basis of the vector space ${\cal V}$. 
The corresponding tangent space algebra ${\cal A}\ $ is spanned by 
components $\{ X(s,s_3\,;{\dot{s}},{\dot{s}}_3)\} $ of vector fields 
$\{X(s,{\dot{s}})\,;\,(s,{\dot{s}})\in\Lambda_p\}$ in 
one-to-one correspondence with the coordinate tensors. Denoting
$$
S=(s,{\dot{s}})\  ,\quad \overline{S}=\{s,s_3\,;{\dot{s}},{\dot{s}}_3\}\ ,
\quad
R=(r,{\dot{r}})\ ,\quad \overline{R}=\{r,r_3\,;{\dot{r}},\dot r_3\}\ ,\quad
\mbox{etc.},
$$
and defining the sign of the graded bracket as
$$
S\bullet R=R\bullet S =(-1)^{4(s+{\dot{s}})(r+{\dot{r}})+1}\  ,
$$
the most general Lorentz covariant superalgebra ${\cal A}\ $ of vector fields 
takes the form
\begin{eqnarray}
&& \left [\ X(\overline{S})\ ,\
     X(\overline{R})\ \right ]_{S\bullet R}  
\nonumber \\[8pt]
     &&=\quad \sum_{(v,{\dot{v}})\in\Gamma(S,R)\cap\Lambda_p}
  C(s,s_3,r,r_3\,;v,s_3+r_3)\ 
  C({\dot{s}},{\dot{s}}_3,{\dot{r}},\dot r_3\,;{\dot{v}},{\dot{s}}_3+\dot r_3)
\nonumber \\
     &&\hspace{3.5truecm}\times\, t(s,{\dot{s}}\,,r,{\dot{r}}\,,v,{\dot{v}})\ 
     X(v,s_3+r_3\,;{\dot{v}},{\dot{s}}_3+\dot r_3)\quad .
\label{comXX}
\end{eqnarray} 
Here  $C(s,s_3,r,r_3\,;v,s_3+r_3)$ are $SU(2)$ Clebsch-Gordan coefficients and
$\Gamma(S,R)$ denotes the indices in the double Clebsch-Gordon decomposition: 
$$
\Gamma(S,R) = \{\ (v,{\dot{v}})\ ;\quad v \in \gamma(s,r)\ ,\   
\ \dot v \in\ \gamma({\dot{s}},{\dot{r}}) \}\ ;\quad
\gamma(s,r) = \{\ s{+}r\,,\ s{+}r{-}1\,,\ \dots,\ |s-r|\ \}\ .
$$
The $t(s,{\dot{s}}\,,r,{\dot{r}}\,,v,{\dot{v}})$'s are structure constants. 
Subject to superskewsymmetry requirements inherited from the graded bracket 
and quadratic constraints implied by the super Jacobi identities, they 
parametrise the moduli space of superalgebras with relations (\ref{comXX}).

The vector fields $\,X\in {\cal A}{=}{\cal A}_0{+}{\cal A}_1\ $ act as 
superderivations on functions of the $Y$'s. We require that the action of 
${\cal A}$ on ${\cal V}$ corresponds
to a linear transformation. The combined vector space thus has
supercommutation relations of a generalised Heisenberg superalgebra, with 
the most general Lorentz covariant relations between the $X$'s
and the $Y$'s taking the form
\begin{eqnarray}
&&\left [\ X(\overline{S}),
Y(\overline{R})\ \right ]_{S\bullet R}  
\nonumber \\[8pt]
&&=\quad \sum_{(v,{\dot{v}})\in\Gamma(S,R)\cap\Lambda_p}
\quad    C(s,s_3,r,r_3\,;v,s_3+r_3)\,
 C({\dot{s}},{\dot{s}}_3,{\dot{r}},\dot r_3\,;{\dot{v}},{\dot{s}}_3+\dot r_3)
\nonumber \\
 &&\hspace{3.9truecm} \times\,  u(s,{\dot{s}}\,,r,{\dot{r}}\,,v,{\dot{v}})\
     Y(v,s_3+r_3\,;{\dot{v}},{\dot{s}}_3+\dot r_3)
\nonumber \\[8pt]
&&\quad +\quad   C(s,s_3,s,-s_3\,;0,0)\
           C({\dot{s}},{\dot{s}}_3,{\dot{s}},-{\dot{s}}_3\,;0,0)\
      c(s,{\dot{s}})\ \delta_{sr}\,\delta_{{\dot{s}}{\dot{r}}}\,
      \delta_{s_3+r_3,0}\, \delta_{{\dot{s}}_3+\dot r_3,0}
\ \ \ .
\label{comXY}
\end{eqnarray} 
The $u$'s are further structure constants and the `central' parameters
$c(s,{\dot{s}})$ determine a bilinear pairing between the $X$'s and the $Y$'s,
$$
 <.\,,\,.>\, :\, {\cal A} \times {\cal V} 
\rightarrow {\cal C}\ =\ \{ c(s,{\dot{s}})\ ;\ (s,{\dot{s}})\in\Lambda_p \}
$$
given by
$$
<X(\overline{S}), Y(\overline{R})>
\ =\ 
c(s,{\dot{s}})\
C(s,s_3,s,-s_3\,;0,0)\
           C({\dot{s}},{\dot{s}}_3,{\dot{s}},-{\dot{s}}_3\,;0,0)\
 \delta_{sr}\,\delta_{{\dot{s}}{\dot{r}}}\, \delta_{s_3+r_3,0}\,
\delta_{{\dot{s}}_3+\dot r_3,0}\quad , 
$$
where the CG coefficients $C(s,s_3,s,-s_3\,;0,0)\ $ denote Wigner's 
`metric' invariant in the representation space of fixed spin $s$.

Therefore, requiring Lorentz covariance determines the space of a priori 
allowed structure constants $\{t,u,c\}$. The combined vector space 
$\,{\cal G}{=}{\cal A}{+}{\cal V}\,$ forms a superalgebra if these structure 
constants are subject to quadratic equations, the satisfaction of which make
the super Jacobi identities amongst the $X$'s and the $Y$'s automatic.
These quadratic equations define the moduli space of spin $p$ Lorentz covariant
Heisenberg superalgebras. For any fixed set $\Lambda_p$, these equations allow 
determination using explicit values for Clebsch-Gordan coefficients and $6j$
symbols (from e.g. \cite{DN3}). Particular solutions of these equations
provide concrete examples of Lorentz covariant spin $p$ superspaces; and
they exist for any finite or infinite $p$. 

\section{Spin 2 superspaces}
The hyperspaces ${\cal M}$\ serve as models for configuration spaces, or for
moduli spaces of solutions, of Lorentz invariant field theories; and the
supercommutations relations for ${\cal G}$ provide canonical supercommutation
relations for the corresponding phase spaces.  They therefore provide an
algebraic description of the local symplectic structure.  With this
application in mind, it is clear that algebras including spins up to two are
of possible significance for the canonical quantisation of gravity and
supergravity theories.  By including generators transforming according to
every Lorentz representation (with unit multiplicity) having spin up to two,
the complete set of quadratic equations for the Lorentz covariant structure
constants were determined in \cite{DN3}.  These defining equations for {\it
spin two Heisenberg superalgebras} are highly overdetermined.  Nevertheless,
non-trivial solutions can indeed be found.  Explicit examples of algebras
${\cal G}$ were presented for spins $s{+}{\dot{s}}$ up to 
${\textstyle{3\over 2}}$ in \cite{DN1,DN2}; and for spins up to 2 in 
\cite{DN3}. The latter have a
representation content reminiscent of simple (super)gravity theories:  a
metric represented by canonically conjugate variables $X,Y$ transforming as
$(0,0)\oplus(1,1)$ coupled to single copies of other (spin $<2$)
representations, including Rarita-Schwinger representations
$(1,{\textstyle{1\over 2}})\oplus({\textstyle{1\over 2}},1)$ corresponding to
gravitino phase-space variables.  Concrete physical application to the
canonical quantisation of supergravity theories, however, remains for future
investigation.

\section{Generalised self-duality}
A further application of these hyperspaces considers gauge fields on 
${\cal M}$\  \cite{DN1,DN2} :  
Associating a gauge potential $A$ (depending on the $Y$'s) to each of these 
generalised derivatives, we define gauge-covariant derivatives
$\  D(s,{\dot{s}}){=}X(s,{\dot{s}}){+}A(s,{\dot{s}})\,$. 
The curvature components $F(s,s_3\,;{\dot{s}},{\dot{s}}_3)$ 
which are a priori non-zero are in turn defined in a consistent fashion by:
\begin{eqnarray}
&& \left [\ D(\overline{S})\ ,\
     D(\overline{R})\ \right ]_{S\bullet R}  
\nonumber \\[8pt]
     &&-\quad \sum_{(v,{\dot{v}})\in\widehat\Gamma(S,R)\cap\Lambda_p}
 C(s,s_3,r,r_3\,;v,s_3+r_3)\ 
 C({\dot{s}},{\dot{s}}_3,{\dot{r}},\dot r_3\,;{\dot{v}},{\dot{s}}_3+\dot r_3)
\nonumber \\
&&\hspace{3.5truecm}\times\, t(s,{\dot{s}}\,,r,{\dot{r}}\,,v,{\dot{v}})\ 
     D(v,s_3+r_3\,;{\dot{v}},{\dot{s}}_3+\dot r_3)
 \nonumber \\[8pt]
     &&=\quad \sum_{(v,{\dot{v}})\in\widehat\Gamma(S,R)} 
     C(s,s_3,r,r_3\,;v,s_3+r_3)\ 
C({\dot{s}},{\dot{s}}_3,{\dot{r}},\dot r_3\,;{\dot{v}},{\dot{s}}_3+\dot r_3)
\nonumber \\
     &&\hspace{3.5truecm}\times\,  
     F(v,s_3+r_3\,;{\dot{v}},{\dot{s}}_3+\dot r_3)    
     \quad ,
\label{comDD}
\end{eqnarray} 
where the sums denote Clebsch-Gordon series restricted to terms having the
superskewsymmetry of the graded bracket by choosing
$$
\widehat\Gamma(S,R) = 
\{ 
(v,{\dot{v}})\in\Gamma(S,R)\ |\  v+{\dot{v}} = 
(s+{\dot{s}})+(r+{\dot{r}})-4(s{+}{\dot{s}})(r{+}{\dot{r}})-1\ \mbox{mod}\ 2
\}\ .
$$
Curvature components thus defined form curvature tensors $F(S,R)$
for all pairs $S,R\in\Lambda_p$.

Generalised self-duality \cite{DN1,DN2} corresponds to setting the
following curvature representations to zero,
$$
F(v,{\dot{s}}{+}{\dot{r}})\ =\ 0\quad 
\mbox{for all}\quad  v\in\gamma(s,r)\quad
\mbox{and}\quad  (s,{\dot{s}}),(r,{\dot{r}})\in\Lambda_p\ .
$$
Similarly, generalised anti-self-duality corresponds to the imposition of the 
constraints
$$
F(s{+}v,{\dot{v}})\ =\ 0\quad 
\mbox{for all}\quad {\dot{v}}\in\gamma({\dot{s}},{\dot{r}})\quad
\mbox{and}\quad  (s,{\dot{s}}),(r,{\dot{r}})\in\Lambda_p\ .
$$
A third class of interesting constraints are the {\it light-like integrable 
systems}
$$
F(s{+}r,{\dot{s}}{+}{\dot{r}})\ =\ 0\quad 
\mbox{for all}\quad  (s,{\dot{s}}),(r,{\dot{r}})\in\Lambda_p\ .
$$
All three classes are $SO(4,{\bubble C})$-covariant equations. They are, 
moreover, amenable to generalised twistor-like transforms (for example, on 
the lines of the discussion in \cite{DO}). These systems therefore provide 
an infinitely large hierarchy of gauge- and Lorentz-covariant solvable 
systems. The corresponding linear (Lax) systems are given in \cite{DN2}.

For example, the simple super-Poincar\'e set, 
$\,\Lambda_1 = \{({\textstyle{1\over 2}},{\textstyle{1\over 2}}), 
({\textstyle{1\over 2}},0), (0,{\textstyle{1\over 2}})\} 
\equiv \{V,S,\bar S\}$,
has the set of a priori non-zero curvature tensors,
$$
\begin{array}{rll}
\widehat\Gamma(V,V) 
&\Leftrightarrow& 
\{ F_{VV}(0,1), F_{VV}(1,0) \}
\\[5pt]
\widehat\Gamma(V,S) 
&\Leftrightarrow& 
\{ F_{VS}(1,{\textstyle{1\over 2}}), F_{VS}(0,{\textstyle{1\over 2}}) \}
\\[5pt]
\widehat\Gamma(V,\bar S) 
&\Leftrightarrow& 
\{ F_{V\bar S}({\textstyle{1\over 2}},1) , 
   F_{V\bar S}({\textstyle{1\over 2}},0) \}
\\[5pt]
\widehat\Gamma(S,S) 
&\Leftrightarrow& 
\{ F_{SS}(1,0) \}
\\[5pt]
\widehat\Gamma(S,\bar S) 
&\Leftrightarrow& 
\{ F_{S\bar S}({\textstyle{1\over 2}},{\textstyle{1\over 2}}) \}
\\[5pt]
\widehat\Gamma(\bar S,\bar S) 
&\Leftrightarrow& 
\{ F_{\bar S\bar S}(0,1) \}
\end{array}
$$
For this choice, the above three sets of zero curvature tensors correspond to
the following:
\begin{center}
\begin{tabular}{|c  |c   |c |}
\hline
$F(v,{\dot{s}}{+}{\dot{r}})$                  
&$F(s{+}v,{\dot{v}})$  
&$F(s{+}r,{\dot{s}}{+}{\dot{r}})$
\\[5pt]
\hline
$F_{VV}(0,1)$                     
&$F_{VV}(1,0)$    
& 
\\[5pt]
$F_{VS}(1,{\textstyle{1\over 2}}), F_{VS}(0,{\textstyle{1\over 2}})$
&$F_{VS}(1,{\textstyle{1\over 2}})$
&$F_{VS}(1,{\textstyle{1\over 2}})$
\\[5pt]
$F_{V\bar S}({\textstyle{1\over 2}},1)$
&$F_{V\bar S}({\textstyle{1\over 2}},1),F_{V\bar S}({\textstyle{1\over 2}},0)$
&$F_{V\bar S}({\textstyle{1\over 2}},1)$
\\[5pt]
$F_{SS}(1,0)$                     
&$F_{SS}(1,0)$   
&$F_{SS}(1,0)$
\\[5pt]
$F_{S\bar S}({\textstyle{1\over 2}},{\textstyle{1\over 2}})$    
&$F_{S\bar S}({\textstyle{1\over 2}},{\textstyle{1\over 2}})$
&$F_{S\bar S}({\textstyle{1\over 2}},{\textstyle{1\over 2}})$
\\[5pt]   
$F_{\bar S\bar S}(0,1)$ 
&$F_{\bar S\bar S}(0,1)$
&$F_{\bar S\bar S}(0,1)$
\\
\hline
\end{tabular}                                          
\end{center}

\noindent
These correspond respectively to the supersymmetrisation of the standard 
four-dimensional self-duality condition, $F_{VV}(0,1)=0$, the 
supersymmetrisation of the anti-self-duality condition, $F_{VV}(1,0)=0$, 
and the `conventional constraints' of $N{=}1$ supersymmetric gauge theory. 
Further examples are discussed in \cite{DN2}, including the celebrated $N{=}3$
constraints which are equivalent to the full equations of motion.

\section{Concluding remarks}
Although we remain in the realm of supercommutative geometry, with
$\ [{\cal V},{\cal V}]_{\bullet}{=}0\,$, a generalisation to 
non-supercommutative geometry is  clearly a further possibility, with the 
simplest superalgebra variant having 
$\ [\,.\,,\,.\,] : {\cal V} \times {\cal V} \rightarrow {\cal V}\ $ 
such that 
$\ [{\cal V}_\alpha ,{\cal V}_\beta]_{\bullet} 
\subset {\cal V}_{\alpha+\beta}\ $. 
Further generalisations, replacing this superalgebra structure, for instance,
by $q$-deformed supercommutation relations, may also be considered along 
the lines of the present investigation. 

We have considered an element of ${\cal A}, {\cal V}$ to be of bosonic type if
its spin $(s{+}{\dot{s}})$ is an integer and of fermionic type if its spin is
a genuine half-integer; and we have assumed the corresponding statistics.
We note, however, that this assignment is a purely
conventional one, motivated by the spin-statistics theorem.  This can indeed
be lifted, if required, to yield Lie algebra (rather than superalgebra)
extensions of the Poincar\'e algebra containing integer and half-integer spin
elements, all of even statistics. Such algebras maintain, nevertheless, their
${\bubble Z}_2$-graded nature \cite{AC}. Such a variant of the supersymmetry
algebra was recently shown to be the target space symmetry of the $N{=}2$
string \cite{DL} and the space of string physical states was shown to be
elegantly and compactly describable in terms of a self-dual field on a 
hyperspace with a vectorial and an even-spinorial coordinate. On such 
hyperspaces with even spinorial coordinates, the generalised self-duality 
equations are the same as those given above; the only difference being that 
the a priori set of curvature components is determined from commutators, rather
than supercommutators, between covariant derivatives.


\goodbreak
\end{document}